\documentclass[prr,aps,twocolumn,
preprintnumbers,superscriptaddress,longbibliography]{revtex4-2}
\usepackage{amsmath,amssymb,amsfonts,bm}
\usepackage{mathrsfs}
\usepackage{braket}
\usepackage{graphicx}
\usepackage{subcaption}
\usepackage{mathtools}
\usepackage{color}
\usepackage{caption}
\usepackage{xcolor}
\usepackage[colorlinks,linkcolor=blue,citecolor=blue]{hyperref}
\usepackage{comment}
\usepackage{physics}
\hypersetup{
	colorlinks=true, 
	linkcolor=violet,  
	urlcolor=blue,    
	anchorcolor=green, 
	citecolor=violet   
}
\definecolor{UTokyoblue}{RGB}{11,135,231}
\definecolor{UTokyoyellow}{RGB}{247,199,0}

\usepackage{tikz}
\usepackage{tikz-layers}
\usepackage{pgfplots}
\usetikzlibrary{matrix, positioning, arrows.meta,math,calc,decorations.pathmorphing}
\tikzset{>=Latex,	UTokyo/.style ={color = UTokyoblue, line width =2pt, rounded corners},
	circlebox/.style ={
		circle , 
		minimum size =0.5cm, 
		draw=violet ,
		fill=blue!50 
	}
	,
	box/.style ={
		rectangle, 
		rounded corners =5pt, 
		minimum width =50pt, 
		minimum height =20pt, 
		inner sep=5pt, 
		draw=black 
	},
}

\newcommand{\T}{\mbox{\tiny T}}
\newcommand{\B}{\mbox{\tiny B}}
\newcommand{\tot}{\mbox{\tiny tot}}
\newcommand{\F}{\mbox{\tiny F}}
\newcommand{\s}{\mbox{\tiny S}}

\newcommand{\hP}{\hat{P}}
\newcommand{\hQ}{\hat{Q}}
\newcommand{\hp}{\hat{p}}
\newcommand{\hx}{\hat{x}}

\newcommand{\as}{\alpha}
\newcommand{\nl}{\nonumber \\}
\newcommand{\ii}[3]{\int_{#1}^{#2}\mathrm{d}#3\,}

\newcommand{\eq}[1]{Eq.\,(\ref{#1})}
\newcommand{\eqnot}[1]{\,(\ref{#1})}

\newcommand{\eqtwo}[2]{Eqs.\,(\ref{#1}) and (\ref{#2})}

\newcommand{\cf}[1]{[cf.\,Eq.\,(\ref{#1})]}

\newcommand{\citeref}[1]{\,Ref.\,\cite{#1}}

\begin{document}
\title{Quantum Counterpart of Equipartition Theorem: A Möbius Inversion Approach}

\author{Xin-Hai Tong*}
\affiliation{Department of Physics, The University of Tokyo, 5-1-5 Kashiwanoha, Kashiwa-shi, Chiba 277-8574, Japan}
\email[]{xinhai@iis.u-tokyo.ac.jp}
\author{Yao Wang}
\affiliation{Key Laboratory of Precision and Intelligent Chemistry, University of Science and Technology of China,
	Hefei, Anhui 230026, China}

\begin{abstract}
The equipartition theorem is crucial in classical statistical physics, and recent studies have revealed its quantum counterpart for specific systems. This raises the question: does a quantum counterpart of the equipartition theorem exist for any given system, and if so, what is its concrete form? In this Letter, we employ the Möbius inversion approach to address these questions, providing a criterion to determine whether a system adheres to the quantum counterpart of the equipartition theorem. If it does, the corresponding distribution function can be readily derived. Furthermore, we construct the fermionic version of the  criterion in a manner analogous to the bosonic case.
\end{abstract}

\maketitle
\paragraph*{Introduction. ---}
The equipartition theorem, a fundamental law in classical statistical physics, plays a crucial role in understanding the distribution of energy among the different degrees of freedom of a system in thermal equilibrium.
Proposed in the late nineteenth century, the theorem provides a statistical basis for predicting the average energy associated with each degree of freedom in a classical system \cite{schrodinger1989statistical,reif1998fundamentals}. 
It forms a cornerstone in the bridge between the microscopic world of particles and the macroscopic observables of thermodynamics \cite{huang2008statistical}.
The equipartition theorem states that, 
in thermal equilibrium, the energy for the each degree of freedom is simply
\begin{equation}\label{classicaleet}
	\begin{aligned}[b]
	E_{i}(T)=k_{\B}T/2,
	\end{aligned}
\end{equation}
where $k_{\B}$ is the Boltzmann constant and $T$ the temperature.
This theorem proves invaluable in understanding the behavior of gases, solids, and other classical systems, forming a foundation for the development of statistical mechanics \cite{Dattagupta2010, Campisi2007, Koide2011, Abreu2020, Abreu2020a, Barboza2015, Ziel1973, Sarpeshkar1993, Matheny2013}.

Formally, by setting $\beta\equiv1/(k_{\B}T)$, the inverse temperature, we may recast the classical equipartition theorem \eq{classicaleet} as
\begin{align}\label{eet}
	E_{i}(\beta)=\mathbb{E}_{i}[\mathcal{E}(\omega,\beta)]\coloneq\ii{0}{\infty}{\omega}\mathbb{P}_{i}(\omega)\mathcal{E}(\omega,\beta),
\end{align}
where  $E_{i}$, the mean energy contributed by the $i$th degree of freedom, is expressed as the expectation ($\mathbb{E}_{i}[\bullet]$) of the energy density $\mathcal{E}(\omega,\beta)$ with respect to the distribution $\mathbb{P}_{i}(\omega)$.
In classical scenario,  $\mathcal{E}(\omega,\beta)=1/(2\beta)$, which is independent of $\omega$. This together with the normalization condition, $\ii{0}{\infty}{\omega}\mathbb{P}_{i}(\omega)=1$, which recovers \eq{classicaleet}.

Recently, many researchers \cite{spiechowicz2018quantum,Bialas2018,spiechowicz2019superstatistics,Bialas2019,Kaur2022,spiechowicz2021energy,Kaur2022a,Ghosh_2023,kaur2023partition,Tong2024} try to extend the classical equipartition theorem to the quantum regime with several models, such as the electrical circuits \cite{Ghosh_2023}, the Brownian oscillators \cite{Bialas2018,Ghosh2023,Kaur2021,spiechowicz2018quantum},  dissipative diamagnetism \cite{Bialas2018,spiechowicz2019superstatistics} and considering kinetic energy for
a more general setup \cite{Luczka2020}. 
The quantum counterpart of equipartition theorem also acquires the form of \eq{eet}, but now the energy density $\mathcal{E}(\omega,\beta)$ generally depends on $\omega$. Though the energy of different degree of freedom $i$ differs from each other, the enenrgy density $\mathcal{E}\qty(\omega,\beta)$ is universal for all the
degrees of freedom, representing the “equipartition” in quantum
sense \cite{Tong2024}.
In these researches, the systems are assumed to be quadratic and 
$\mathcal{E}(\omega,\beta)$ is set to be $(\hbar\omega/4)\coth(\hbar\beta\omega/2)$, which is the energy of the quantum harmonic oscillator in the equilibrium thermal state. This $\mathcal{E}$ can be reduced to the classical case \eq{classicaleet} since $\lim_{\hbar\rightarrow 0}(\hbar\omega/4)\coth(\hbar\beta\omega/2)=1/(2\beta)$,
as explained below \eq{eet}.
The normalized distribution functions $\mathbb{P}_{i}(\omega)$ are also explicitly obtained in these quadratic systems \cite{Bialas2019,Tong2024}. 
Moreover, for the fermionic system, the quantum  counterpart of equipartition theorem is also investigated \cite{Kaur2022}.
They altogether provide novel insights for the accurate and convenient evaluations of thermodynamic quantities \cite{Tong2024,Kaur2022a,Ghosh2023}.

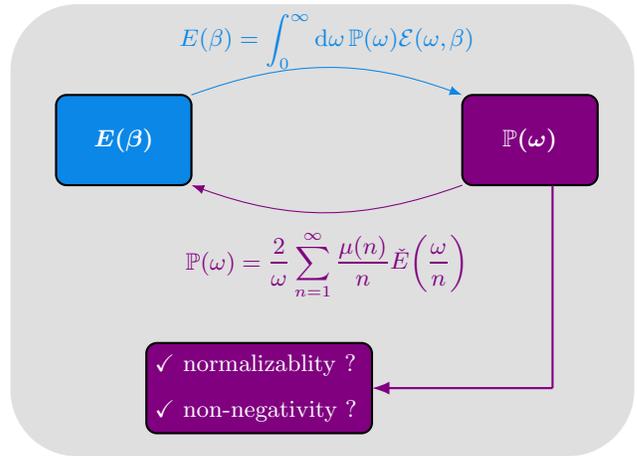
\begin{figure}[h]
	\centering
	\begin{tikzpicture}[shift={(0,0)}]
		\def\a{0.6}
		\def \r{0.075*\a}
		\def\deltax{9}
		\def\deltay{9}
		
	    \draw [ultra thin,rounded corners=25pt,fill=lightgray!50,draw=lightgray!50] (-7*\a,-5*\a) rectangle (6.8*\a,5*\a);
		
		\begin{scope}[shift={(0,-\a)}]
		
		\draw [thick,fill=UTokyoblue,rounded corners]  (-6*\a,2*\a) rectangle 
		+(3*\a,2*\a);
		\node [white] at (-4.5*\a,3*\a) {$\bm{E(\beta)}$};
		
			\draw [thick,fill=violet,rounded corners]  (-6*\a+\deltax*\a,2*\a) rectangle 
		+(3*\a,2*\a);
		\node [white] at (-4.5*\a+\deltax*\a,3*\a) {$\bm{\mathbb{P}(\omega)}$};
		
	\path [->, UTokyoblue] (-3*\a,4*\a) edge  [bend left=20](3*\a,4*\a);
		\path [->, violet] (3*\a,2*\a) edge  [bend left=20](-3*\a,2*\a);
		
		\node [UTokyoblue,above] at (0,4.4*\a) {$\begin{aligned}
			E(\beta)=\ii{0}{\infty}{\omega}\mathbb{P}(\omega)\mathcal{E}(\omega,\beta)
		\end{aligned}$ }   ;	
	
		\node [violet,below] at (0,1.2*\a) {$\begin{aligned}
		\mathbb{P}(\omega)
		=\frac{2}{\omega}\sum_{n=1}^{\infty}\frac{\mu(n)}{n}
		\check{E}\bigg(\frac{\omega}{n}\bigg)
		\end{aligned}$ }   ;

\end{scope}
	\draw [thick,violet] (5*\a,\a)--(5*\a,-3.5*\a) ;
	\draw [->,thick,violet] (5*\a,-3.5*\a)--(\a,-3.5*\a);
	\draw [thick,rounded corners,fill=violet] (-4*\a,-4.5*\a) rectangle (\a,-2.5*\a);
	\node [white,right]at (-4*\a,-3*\a) {$\checkmark$ normalizablity ?};
		\node [white,right]at (-4*\a,-4*\a) {$\checkmark$ non-negativity ?};
	\end{tikzpicture}
		\captionsetup{justification=raggedright,singlelinecheck=false}
	\caption{An illustration for the interplay between energy spectrum $E(\beta)$ and distribution funciton $\mathbb{P}(\omega)$ with the help of \eqtwo{eet}{IEET}. The normalizablity and non-negativity  should also be checked for $\mathbb{P}(\omega)$. }
	\label{1}
\end{figure}

For more general systems beyond above mentioned quadratic models, does the  quantum  counterpart of equipartition theorem still hold? 
If so, how to obtain  the corresponding distribution function $\mathbb{P}_{i}(\omega)$?
Answers to these questions will serve as a promising methodology for studying the quantum thermodynamics.
This Letter aims to
use a universal approach, the Möbius inversion, to answer these questions.
It originates from the number theory \cite{hardy1979introduction} and has been used in various  inverse problems in physics \cite{chen1990modified,nan1998unified,chen1992carlsson}. 
Based on the Möbius inversion,  we  give a criterion to determine whether the  quantum  counterpart of equipartition theorem holds for a given system. 
Furthermore, it tells us 
how to obtain the distribution $\mathbb{P}_{i}(\omega)$ with this systematic approach.  
We implement the proposed formulas to some typical models, including the free photon gas \cite{huang2009introduction,landau2013statistical},  the harmonic oscillator \cite{griffiths2018introduction}, the Riemann gas \cite{schumayer2011colloquium,julia1990statistical}, and the Ising model \cite{brush1967history, huang2009introduction}.
It is worth noting that this method applies to both bosonic and fermionic scenarios.
%
%
%
%


\paragraph*{Quantum counterpart of equipartition theorem for quadratic systems.---}
As a prelude, we first briefly illustrate the quantum counterpart of equipartition theorem with an example of harmonic oscillator \cite{Tong2024,Bialas2019,Ghosh_2023,Kaur2021,Kaur2022a,Bialas2018,spiechowicz2018quantum,spiechowicz2019superstatistics,spiechowicz2021energy,Ghosh2023,kaur2023partition}.
The quantum counterpart of equipartition theorem was discussed in the scenario of open systems, whose simplest quadratic model reads
\begin{align}\label{2}
	H_{\T}=H_{\s}+\sum_{j}\left[\frac{\hp_{j}^{2}}{2m_{j}}
	+\frac{1}{2}m_{j}\omega_{j}^{2}\bigg(\hx_{j}-\frac{c_{j}\hat{Q}}{m_{j}\omega_{j}^{2}}\bigg)^{2}\right]
\end{align}
with
\begin{align}\label{3}
	H_{\s}=\frac{\hP^{2}}{2M}+\frac{1}{2}M\Omega^{2}\hQ^{2}.
\end{align}
This is the Calderia-Leggett model \cite{caldeira1981influence} describing a harmonic oscillator ($\hQ, \hP$) of mass $M$ coupled to  the heat bath ($\{\hx_{j}, \hp_j\}$).
For the system oscillator, the kinetic energy and the potential energy are $E_{\rm k}(\beta)=\langle \hP^{2} \rangle/(2M)$ and  $E_{\rm p}(\beta)=M\Omega^{2}\langle \hQ^{2}\rangle/2$, respectively. The average is
defined over the total Gibbs state, i.e., $ \langle \bullet  \rangle \coloneq \operatorname{tr}[\bullet e^{-\beta H_{\T}}]/\operatorname{tr}e^{-\beta H_{\T}}$.
For brevity, we set $\hbar=1$ hereafter. 
It was shown \cite{Bialas2019,Ghosh2023} that
both $E_{\rm k}(\beta)$ and $E_{\rm p}(\beta)$ 
can be expressed in the form of \eq{eet} with 
\begin{equation}
	\mathcal{E}(\omega,\beta)=\frac{\omega}{4}\coth
	\bigg(\frac{\beta\omega}{2}\bigg)  
\end{equation} 
and
\begin{align}
	\mathbb{P}_{\rm k}(\omega)=\frac{2M\omega}{\pi}\operatorname{Im}J(\omega),\quad \mathbb{P}_{\rm p}(\omega)=\frac{2M\Omega^{2}}{\pi\omega}\operatorname{Im}J(\omega).
\end{align}
Here,  $J(\omega)$ denotes the generalized susceptibility \cite{Ghosh2023}.
It was verified that
$\mathbb{P}_{\mathrm{p,k}}(\omega)$ satistifies the normalized and nonnegative condition \cite{Ghosh2023,Bialas2019}.
In the classical limit $\hbar\rightarrow 0$, we have 
$\mathcal{E}
(\omega,\beta) \rightarrow 1/(2\beta)$, which
gives rise to the classical equipartition theorem \eq{classicaleet}.
In the weak-coupling limit \cite{Ghosh2023}, $c_{j}\rightarrow0$ for all $j$ in \eq{2}, resulting in
$\mathbb{P}_{\rm p,k}(\omega)\rightarrow \delta(\omega-\Omega)$.
Besides, as explained in \,Refs.\,\cite{Tong2024,Kaur2022a}, the free energy $F(\beta)$ is expressed in the same form by simply switching $\mathcal{E}(\omega,\beta)$ into $\mathcal{F}(\omega,\beta)=\ln(2\sinh(\beta\omega/2))/\beta$, which is the average free energy of the oscillator in the canonical ensemble. For more detailed discussions of general quadratic systems, we refer the readers to Ref.\,\cite{Tong2024}.

\paragraph*{Möbius Inversion Approach.---}
To explore the quantum counterpart of equipartition theorem for general systems, 
it is our task to find a 
non-negative and normalizable $\mathbb{P}(\omega)$ for 
each degree of freedom, given 
the energy spectrum $E(\beta)$.
For brevity, we omit the label of degree of freedom hereafter.
It will be shown below that for any given energy specturm $E(\beta)$ from theoretical calculation or experimental measurement, if the quantum counterpart of equipartition theorem is valid,  then we have 
\begin{align}\label{IEET}
	\mathbb{P}(\omega)
	=\frac{2}{\omega}\sum_{n=1}^{\infty}\frac{\mu(n)}{n}
	\check{E}\bigg(\frac{\omega}{n}\bigg).
\end{align}
Here, $\check{f}(\omega)\coloneq\mathcal{L}^{-1}[f(\beta)]$ denotes the inverse Laplace transform of the function $f(\beta)$, and $\mu(n)$ is the celebrated Möbius function \cite{hardy1979introduction}.

Let us look at \eq{IEET} from another angle.  To ensure the validity of the equipartition in the system with the spectrum $E(\beta)$,  we first obtain $\mathbb{P}(\omega)$ from the right-hand side of \eq{IEET}. It is the next task to check its non-negativity.
Furthermore,  the normalizability requires   $\ii{0}{\infty}{\omega}\mathbb{P}(\omega)$  converge to a finite positive number. This global constant, possibly dependent on  the the size of the system,  shall be absorbed into $\mathbb{P}(\omega)$ \cite{Tong2024}.
By substituding the normalized $\mathbb{P}(\omega)$ into \eq{1}, we obtain the quantum counterpart of equipartition theorem.
On the other hand, if the obtained $P(\omega)$ via \eq{IEET} is without non-negativity and normalizablity, we claim that there is no such a quantum counterpart. 
In this sense, \eq{IEET} supplies  a sufficient and necessary condition to ascertain the presence of the quantum counterpart of equipartition theorem. 
If the equipartition holds, \eqtwo{eet}{IEET} further give a concrete expression of $\mathbb{P}(\omega)$.

Now we give a detailed derivation of the \eq{IEET}.
First notice the following expansion:
\begin{align}\label{bexpansion}
	\mathcal{E}(\omega,\beta)=\frac{\omega}{4}\qty(2\sum_{n=1}^{\infty}e^{-n\beta\omega}+1) \quad \text{for } \omega>0.
\end{align}
By substituting it into \eq{1}, we obtain
\begin{subequations}
	\begin{align}
		E(\beta)
		&=\!\sum_{n=1}^{\infty}\ii{0}{\infty}{\omega}\frac{\omega}{2}\mathbb{P}(\omega)e^{-n\beta\omega}\!+\!\frac{1}{4}\!\ii{0}{\infty}{\omega}\omega\mathbb{P}(\omega)\label{9a}	
		\\
		&= \sum_{n=1}^{\infty}\mathcal{L}\qty[\frac{\omega}{2}\mathbb{P}(\omega)](n\beta)\label{9b}	. 
	\end{align}
\end{subequations}
For the second term on the right-hand side of \eq{9a}, we note that $\lim_{\beta\rightarrow\infty}\mathcal{E}(\omega,\beta)=\omega/4$ and  $E(\infty)=\ii{0}{\infty}{\omega}\omega\mathbb{P}(\omega)/4$ \cf{1}.  Therefore, one may absorb this term into the left-hand side of \eq{9a} to redefine the energy specturm as $E(\beta)-E(\infty)$ \cf{9b}.
To proceed, we consult  the modified Möbius inversion formula \cite{Chen1990}: for two functions $f(x)$ and $g(x)$, we have
\begin{align}\label{A5}
	f(x)=\sum_{n=1}^{\infty}g(nx)\Longleftrightarrow g(x)=\sum_{n=1}^{\infty}\mu(n)f(nx).
\end{align}
By noticing that the right-hand side of \eq{9b} is just a function with respect to $n\beta$,  the M\"obius inversion gives
\begin{align}\label{11}
	\mathcal{L}\qty[\frac{\omega}{2}\mathbb{P}(\omega)](\beta)=\sum_{n=1}^{\infty}\mu(n)E(n\beta),
\end{align}
which is equivalent to
\begin{align}\label{A7}
	\mathbb{P}(\omega)=\frac{2}{\omega}\sum_{n=1}^{\infty}\frac{\mu(n)}{n}\mathcal{L}^{-1}[E(\beta)]\qty(\frac{\omega}{n}).
\end{align}
Here, we have used  $\mathcal{L}^{-1}[E(n\beta)](\omega)=\mathcal{L}^{-1}[E(\beta)](\omega/n)/n$. Then we arrive at \eq{IEET}.

\paragraph*{Typical Examples.---}
Let us turn to several examples to illustrate the procedure. 
Generally speaking, the asymptotic behavior of the energy spectrum at infinite temperature ($\beta=0$) plays a crucial role.
Noting that $\mathcal{E}(\omega,\beta)\sim \beta^{-1}$ and $\mathbb{P}(\omega)$ is normalized, we thereby conclude from \eq{1} that $E(\beta)\sim \beta^{-1}$ for any degree of freedom and so is the total energy.
As a result, the quantum counterpart of equipartition theorem does \emph{not} hold in such as the Ising model \cite{brush1967history,huang2009introduction} and the Riemann gas \cite{schumayer2011colloquium,julia1990statistical},  whose energy spectrums converge to a finite real number when $\beta\rightarrow 0$.
The same criterion also rules out the photon gas governed by the well-known Stefan–Boltzmann law in two and three dimensions, whose total energy spectrums asymptotically behave as $\beta^{-3}$ and $\beta^{-4}$, respectively.

For generality, we set $E^{\tot}(\beta)=\ii{0}{\infty}{k}A(k)\beta^{-k}$ with $A(k)$ being an undetermined function. The inverse Laplace transform of $E^{\tot}(\beta)$ reads $	\check{E}^{\tot}(\omega)=\ii{0}{\infty}{k}A(k)\Gamma(k)\omega^{k-1}$. The distribution function is evaluated to be $	\mathbb{P}(\omega)=\ii{0}{\infty}{k}A(k)\Gamma(k)\omega^{k-2}/\zeta(k)$ \cf{IEET}. Here, we used the property of the Möbius function \cite{apostol1998introduction}, $\sum_{n=1}^{\infty}\mu(n)/n^{s}=1/\zeta(s) \text{ for } \operatorname{Re}s>1$, where $\zeta(s)$ is the Riemann zeta function. Therefore, the key point is to exmamine whether this distribution function satisfies the non-negativity and normalizability, which solely depends on the concrete form of $A(k)$.  As a simple example, we choose $A(k)=C\delta(k-k_{0})$ with a constant $C\in\mathbb{R}^{+}$. The resulted distribution function is $\mathbb{P}(\omega)=C \Gamma(k_{0})\omega^{k_{0}-2}/\zeta(k_{0})$, which cannot be normalized for $k_{0}=3,4$.  This conclusion aligns with our previous analysis of photon gas.

We should emphasize here that the absence of quantum counterpart of equipartition theorem does not show any violation to the fundamental principles of quantum mechanics or statistical physics. One of the prerequisites is the energy specturm behaving asymptotically as $\beta^{-1}$ at high temperature, as discussed at the beginning of this paragraph. Therefore, not all real equilibrium necessarily should adhere to quantum counterpart of equipartition theorem.

Now turn to the linear superposition property. Assume that we have a set of energy spectrums $\qty{E_{i}\qty(\beta)}$, and all of which follow the quantum counterpart of equipartition theorem. We denote the corresponding distribution function as $\qty{\mathbb{P}_{i}\qty(\omega)}$.
Due to the linear property of inverse Laplace transform, the energy spectrum $E\qty(\beta)=\sum_{i}\as_{i}E_{i}\qty(\beta)$ also satisfies the
quantum counterpart of equipartition theorem with the distribution function $\mathbb{P}\qty(\omega)=\sum_{i}\as_{i}\mathbb{P}_{i}\qty(\omega)$, as long as all the coefficients $\qty{\as_{i}}$ are non-negative. This distribution function shall be further normalized as $\mathbb{P}\qty(\omega)=\sum_{i}\as_{i}\mathbb{P}_{i}\qty(\omega)/\sum_{i}\as_{i}$. 
The present results are immediately followed by an example in which
 the energy spectrums are set to be $\{E_{l}\qty(\beta)=\omega_{0}e^{-l\beta\omega_{0}}/\qty(e^{\beta\omega_{0}}-1)\}$ with $\omega_{0}$ a positive constant and $l$ an integer. In this case, we have
\begin{align}\label{232}
	\check{E}_{l}\qty(\omega)&=\omega_{0}\mathcal{L}^{-1}\qty[\frac{e^{-\qty(l+1)\beta\omega_{0}}}{1-e^{-\beta\omega_{0}}}]=\omega_{0}\mathcal{L}^{-1}\qty[\sum_{n=l+1}^{\infty}e^{-n\beta\omega_{0}}]
	\nl &=\omega_{0}\sum_{n=l+1}^{\infty}\delta(\omega-n\omega_{0}).
\end{align}
From \eq{232} and Möbius inversion, we obtain the corresponding distribution function,
\begin{align}\label{233}
	\mathbb{P}_{l}\qty(\omega)&=\frac{2}{\omega}\sum_{n=l+1}^{\infty}\frac{\mu(n)}{n}\omega_{0}\sum_{m=1}^{\infty}\delta(\omega/n-m\omega_{0})
	\nl &=2\sum_{k=l+1}^{\infty}\sum_{n|k}\frac{\mu(n)}{k}\delta\qty(\omega-k\omega_{0})
	\nl &=2\sum_{k=l+1}^{\infty}\frac{\delta_{k,1}}{k}\delta\qty(\omega-k\omega_{0}).
\end{align}
where $n|k$ means  the integer $n$ divides $k$. To obtain the last equality, we have used the identity $\sum_{n|k}\mu(n)=\delta_{k,1}$.
For $l\leq0$, the distribution function \eqnot{233} directly reduces to $2\delta(\omega-\omega_{0})$.
For $l>0$, we have $\mathbb{P}_{l}\qty(\omega)=0$. 
Due to the linear superposition property,  we know that the spectrum 
\begin{align}\label{239}
	E\qty(\beta)=\sum_{l\leq0}\as_{l}E_{l}\qty(\beta)
\end{align}
adheres to quantum counterpart of equipartition theorem  with the distribution function
\begin{align}
	\mathbb{P}\qty(\omega)=2\sum_{l\leq0}\as_{l}\delta(\omega-\omega_{0})
\end{align}
up to a normalization.
Specifically, if we set $\as_{l}=1/4$ for $l=-1,0$ and $\as_{l}=0$ otherwise, \eq{239} reduces to the spectrum of the quantum harmonic oscillator system:
\begin{align}
	E(\beta)=\frac{1}{4}\qty[E_{0}\qty(\beta)+E_{-1}\qty(\beta)]=\frac{\omega_{0}}{4}\coth\frac{\beta\omega_{0}}{2}
\end{align}
with $
\mathbb{P}\qty(\omega)=\delta(\omega-\omega_{0})
$. This result also aligns with our expectation, since for quantum harmonic oscillator we have from \eq{eet} that $(\omega_{0}/4\coth(\beta\omega_{0}/2)=\ii{0}{\infty}{\omega}\mathcal{E}(\omega,\beta)\delta(\omega-\omega_{0})$.

\paragraph*{Fermionic version.---}
Here, we present the fermionic version of the quantum counterpart of the equipartition theorem and its Möbius inverse by analogy.
Note that in the bosonic case, \eq{eet} can be recast as
\begin{align}\label{EET2}
	E(\beta)&=\frac{1}{4}\ii{-\infty}{\infty}{\omega}\mathbb{P}(\omega)\frac{\omega}{e^{\beta\omega}-1}
	\nl &
	= \frac{1}{4}\ii{-\infty}{\infty}{\omega}\mathbb{P}(\omega)\omega\rho^{\B}(\omega)
\end{align}
with the even extension $\mathbb{P}(-\omega)=\mathbb{P}(\omega)$ for $\omega\geq 0$. The factor $\rho^{\B}(\omega)\coloneq 1/(e^{\beta\omega}-1)$ is recognized as  the expected number of bosonic particles with the energy $\omega$.
In the fermionic case, we just replace $\rho^{\B}(\omega)$ by $\rho^{\F}(\omega)=1/(e^{\beta\omega}+1)$ and obtain
\begin{equation}\label{19}
	\begin{aligned}[b]
		E^{\F}(\beta)&=	 \frac{1}{4}\ii{-\infty}{\infty}{\omega}\mathbb{P}^{\F}(\omega)\omega\rho^{\F}(\omega)
		\\ &=\ii{0}{\infty}{\omega}\mathcal{E}^{\F}(\omega,\beta)\mathbb{P}^{\F}(\omega).
	\end{aligned}
\end{equation}
In the second equality,
we have defined $\mathcal{E}^{\F}(\omega,\beta)\coloneq -(\omega/4)\tanh (\beta\omega/2)$ and set $\mathbb{P}^{\F}(-\omega)=\mathbb{P}^{\F}(\omega)$.
This result is equivalent to that in \citeref{Kaur2022}. 

To utilize Möbius inversion, we follow a similiar procedures as in the bosonic case. Firstly, we substitute the following series expansion, 
\begin{align}
	\mathcal{E}^{\F}(\omega,\beta)=-\frac{\omega}{4}\bigg[2\sum_{n=1}^{\infty}(-1)^{n}e^{-n\beta\omega}+1\bigg],
\end{align}
into the fermionic version \eq{19}, obtaining
\begin{align}\label{A4}
	E(\beta)&=\sum_{n=1}^{\infty}\ii{0}{\infty}{\omega}\frac{\omega}{2}\mathbb{P}(\omega)(-1)^{n-1}e^{-n\beta\omega}
	\nl & \quad 
	-\frac{1}{4}\ii{0}{\infty}{\omega}\omega\mathbb{P}(\omega).
\end{align}
Since $\lim_{\beta\rightarrow\infty}\mathcal{E}^{\F}(\omega,\beta)=-\omega/4$, the second term in \eq{A4} is equal to $E(\infty)$, which can also be absorbed into $E(\beta)$ to redefine the energy specturm. 
We have the following  modified Möbius inversion formula  for alternating series \cite{chen1991inverse}:
\begin{align}\label{22}
	& g(x)=\sum_{n=1}^{\infty}(-1)^{n-1}f(nx) 
	\nl &\Leftrightarrow f(x)=\sum_{n=1}^{\infty}\mu(n)\bigg[\sum_{m=1}^{\infty}2^{m-1}g(2^{m-1}nx) \bigg].
\end{align}
Applying \eq{22} to \eq{A4}, we finally arrive at
\begin{align}\label{fIEET}
	\mathbb{P}(\omega)
	=\frac{2}{\omega}\sum_{n=1}^{\infty}\frac{\mu(n)}{n}
	\sum_{m=1}^{\infty}\check{E}(\frac{\omega}{2^{m-1}n}),
\end{align}
which is the desired fermionic version of the Möbius inversion.
\paragraph*{Summary.---}
In conclusion, we introduced the Möbius inversion to study the existence of a quantum counterpart to the equipartition theorem and derived the distribution function $\mathbb{P}(\omega)$ for a given system. Our approach has been applied to various systems, extending the analysis from bosons to fermions. Future work will explore additional connections between number theory and statistical physics, investigate nontrivial energy spectra in open quantum systems, and examine links between the quantum counterpart of the equipartition theorem and level statistics or random matrix theory \cite{schumayer2011colloquium, haake1991quantum, schomerus2007random}.

\vspace{1em}
We thank Prof.\,Naomichi Hatano for his valuable suggestions. We thank Dr. Jaeha Lee, Ao Yuan and Bufan Zheng for friutful discussions. 
Xinhai Tong was supported by FoPM, WINGS Program, the University of Tokyo. This research was supported by Forefront Physics and Mathematics Program to Drive Transformation (FoPM), a World-Leading Innovative Graduate Study (WINGS) Program, the University of Tokyo. Yao Wang was supported by the National Natural Science Foundation of China (Grant Nos. 22103073 and  22373091).

\end{document}